\documentclass[11pt,twoside]{article}

\usepackage{asp2008n}
\usepackage{times}
\usepackage{epsf}
\usepackage{lscape}

\usepackage{graphicx}
\usepackage{amssymb}
\usepackage{longtable}
\usepackage[figuresright]{rotating}
\usepackage{multirow}

\begin{document}
\title{The Lacerta OB1 Association}   
\author{W. P. Chen$^{1,2}$ \ and H. T. Lee$^{1,3}$ }   
\affil{$^1$Institute of Astronomy, National Central University, 300 Jhongda Road, Jhongli 32054, Taiwan}    
\affil{$^2$Department of Physics, National Central University, 300 Jhongda Road, Jhongli 32001, Taiwan}
\affil{$^3$Institute of Astronomy and Astrophysics, Academia Sinica, P.O. Box 23--141, Taipei 10617, Taiwan}


\begin{abstract} 
Lac\,OB1 is a nearby OB association in its final stage of star formation.  While the member stars 
suggest an expansion time scale of tens of Myr, the latest star formation episode, 
as manifested by the existence of massive and pre-main sequence stars, took place no more than 
a few Myr ago.  The remnant molecular clouds in the region provide evidence of starbirth  
triggered by massive stars.    
\end{abstract}

\section{Introduction } 

The Lacerta OB1 (I Lacertae) association was discovered by
\citet{bla53} as an aggregate of dispersing early-type stars, with an
expansion time scale of a couple million years.  In the review article 
on nearby O associations, \citet{bla64}
listed a distance of 600~pc for Lac\,OB1, estimated 
by means of H$\beta$ photometry by \citet{cra61}, which had quite a
large uncertainty because of the presence of pre-main sequence (PMS)
objects, and a possible age spread among member stars.  A relatively
recent distance determination by \citet{dez99}, derived from the
$Hipparcos$ data, yielded an average distance of $\sim370$~pc.  A
noticeable distance range is obviously expected for a nearby
association, which by itself has a typical size extent of a few
hundred parsecs.  With a distance less than 400~pc,
Lac\,OB1 ranks among the nearest OB associations in the solar
neighborhood, and forms a part of the Gould belt system.   
The interstellar matter associated with the Gould belt is organized 
into a giant expanding ring, called the Lindblad ring \citep{lin73}, 
in whose periphery lie the local stellar associations, including 
Lac\,OB1 \citep{ola82}.  Early radio observations of Lac\,OB1 in the 21~cm 
line of neutral hydrogen were carried out by \citet{rai57}, \citet{how58}, 
and \citet{die60}.   

\citet{bla58} divided Lac\,OB1 into two subgroups, Lac\,OB1a and Lac\,OB1b, on the
basis of stellar proper motions and radial velocities.  The entire
Lac\,OB1 is centered around RA=22h35m and Decl=$+43\fdg3$, and covers the large sky
region $90\deg \la \ell \la 110\deg$ and $-5\deg \la b \la
-25\deg$ \citep{dez99}.  The subgroup Lac\,OB1b is considered younger
and more concentrated, distributed within a $\sim5\deg$ radius centered 
around $(\ell, b)=(97\fdg0, -15\fdg5)$, whereas the older Lac\,OB1a extends
over the remaining region.   \citet{bla58} listed 15 stars for Lac\,OB1a, and
11 stars for Lac\,OB1b.  The Lac\,OB1b harbors the only O star in the
region, 10\,Lac (O9\,V; HIP\,111841).  \Citet{dez99} identified a total
of 96 $Hipparcos$ members for Lac\,OB1, including 1 O, 35 B, 46 A, 1
F, 8 K, 3 M-type stars, 1 carbon star (HIP\,116681) and 1 star without
spectral information (HIP\,111762).  Table~1 lists these 96 stars
together with their 2MASS JHKs photometry.  The first column is the
$Hipparcos$ number, followed by (2) and (3) the star's coordinates,
(4) the apparent $V$ magnitude, (5) $\bv$ color, (6) parallax and (7) the
proper motions.  Columns (8), (9), and (10) are 2MASS magnitudes.  Column (11) 
gives the spectral type, and the last column (12) provides some information 
gathered from SIMBAD.

\Citet{dez99} gave a comprehensive reference list for Lac\,OB1 on the 
following data: 
(1)~Distance. For instance, \citet{les69} estimated 368~pc and 603~pc,  
while \citet{cra76} obtained 417~pc and 479~pc for the subgroups Lac\,OB1a and Lac\,OB1b, 
respectively.  
(2)~Proper motions, $\mu_\ell \cos b = -2.3 \pm 0.1$~mas~yr$^{-1}$, and 
$ \mu_b) = -3.4 \pm 0.1$~mas~yr$^{-1}$. 
(3)~Radial velocity. \citet{bij81} obtained the peak around $v_{\rm rad}^{\rm LSR} \sim -15$~km~s$^{-1}$, 
whereas the $Hipparcos$ Input Catalogue gave an average of $v_{\rm rad} = -13.3$~km~s$^{-1}$.
(4)~Expansion age (e.g., $\sim 2.5 \pm 0.5$~Myr, \citeauthor{les69} \citeyear{les69}).
(5)~Stellar rotation (e.g., \citeauthor{abt62} \citeyear{abt62}).  
(6)~Photometric (e.g., $uvby$, \citeauthor{cra76} \citeyear{cra76}) and spectroscopic (e.g.,
\citeauthor{coy69} \citeyear{coy69}, \citeauthor{lev76}
\citeyear{lev76}, \citeauthor{gue76} \citeyear{gue76}) studies.  
Also useful is the review by \citet{gar94} on the
physical properties and dynamical evolution of OB associations,
including Lac\,OB1.

\section{Sites of Recent Star Formation in Lac\,OB1} 

\begin{figure}[!ht] 
  \plotfiddle{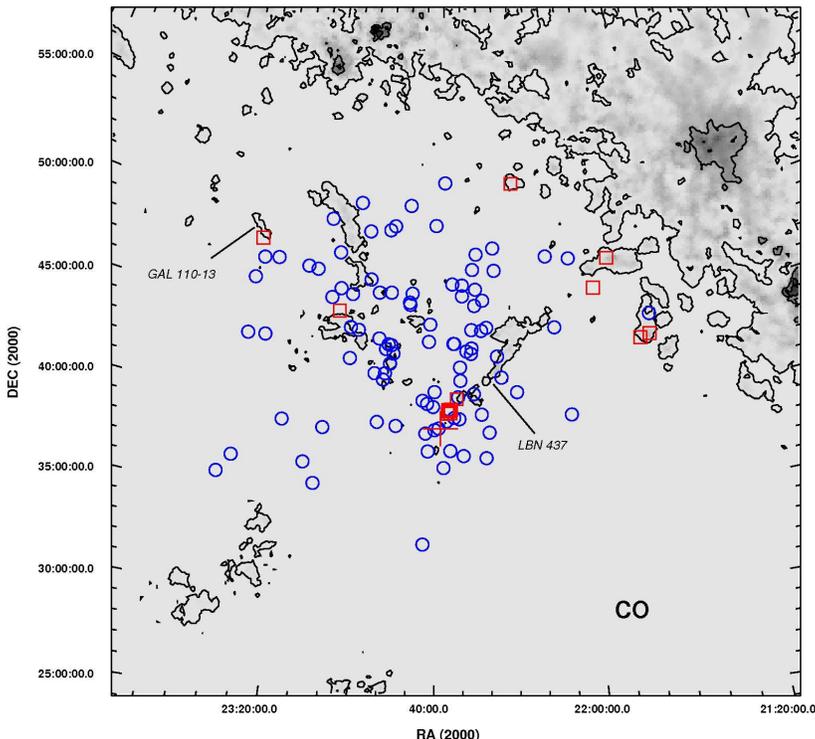}{9cm}{0}{78}{78}{-250}{-30}
 \caption{CO emission in Lac\,OB1 \citep{dam01}.
    The circles mark the positions of the $Hipparcos$ member
    stars \citep{dez99}, and the boxes represent the CTTSs and Herbig
    Ae/Be stars \citep{lee07}.  The O star 10\,Lac is indicated by a
    cross. The Galactic plane is seen on the upper right.  The figure 
    covers roughly the Galactic coordinates from $\ell \sim 75\deg$ to $\sim 120\deg$ 
    and from $b \sim +10\deg$ to $\sim -35\deg$. }
 \label{fig:lacob1co} 
\end{figure}

Despite a considerable number of massive member stars, Lac\,OB1 is
relatively devoid of cloud material.  Two regions---both being remnant
molecular clouds---are known to have had recent star-forming
activities, namely the bright-rimmed cloud LBN\,437 \citep{lyn65} and
the comet-shaped cloud GAL\,110$-$13 \citep{whi49}.   
Figure~\ref{fig:lacob1co} shows the molecular CO emission in the region
\citep{dam01}, along with the $Hipparcos$ members \citep{dez99},
Herbig Ae/Be stars and classical T Tauri stars (CTTSs) \citep{lee07}
in the Lac\,OB1 region.

\subsection{LBN\,437} 

\begin{figure}[!ht] 
  \plotfiddle{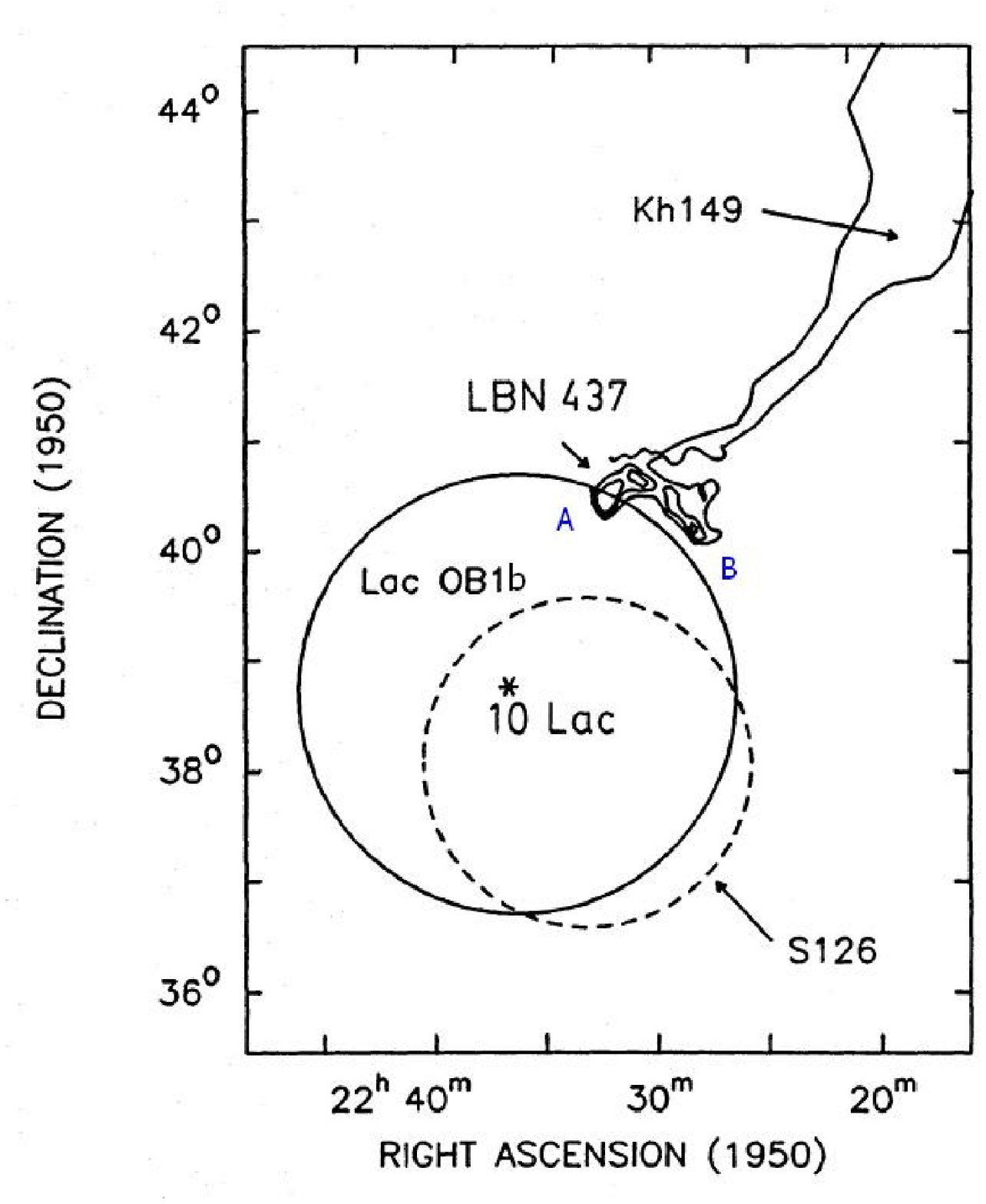}{9cm}{0}{50}{50}{-150}{-80}
 \caption{Schematic of Lac\,OB1 near the the LBN\,437 cloud \citep[modified from][]{ola94}.
        The condensation A and B at the southern end of LBN\,437 are marked.  
        The circle of Lac\,OB1 here refers to the subgroup Lac\,OB1b only.  
     }
 \label{fig:lbn437}
\end{figure}

LBN\,437 is at the edge of an elongated molecular cloud complex Kh\,149 
\citep{kha60}, also known as GAL\,96$-$15 \citep{ode88}, and on the 
border of the H\,II region S\,126 \citep{sha59} excited by
10\,Lac (see Fig.~\ref{fig:lbn437}).  The southern end of LBN\,437  
is forked into two condensations sharing the same mean radial 
velocity.  Condensation A contains a cold, elliptical dense core traced 
by NH$_{3}$ emission, and is associatd with optical reflection nebula and 
luminous young stars, whereas the less massive Condensation B appears 
not associated with any optical stars \citep{ola94}.    

\begin{figure}[!ht] 
   \plotone{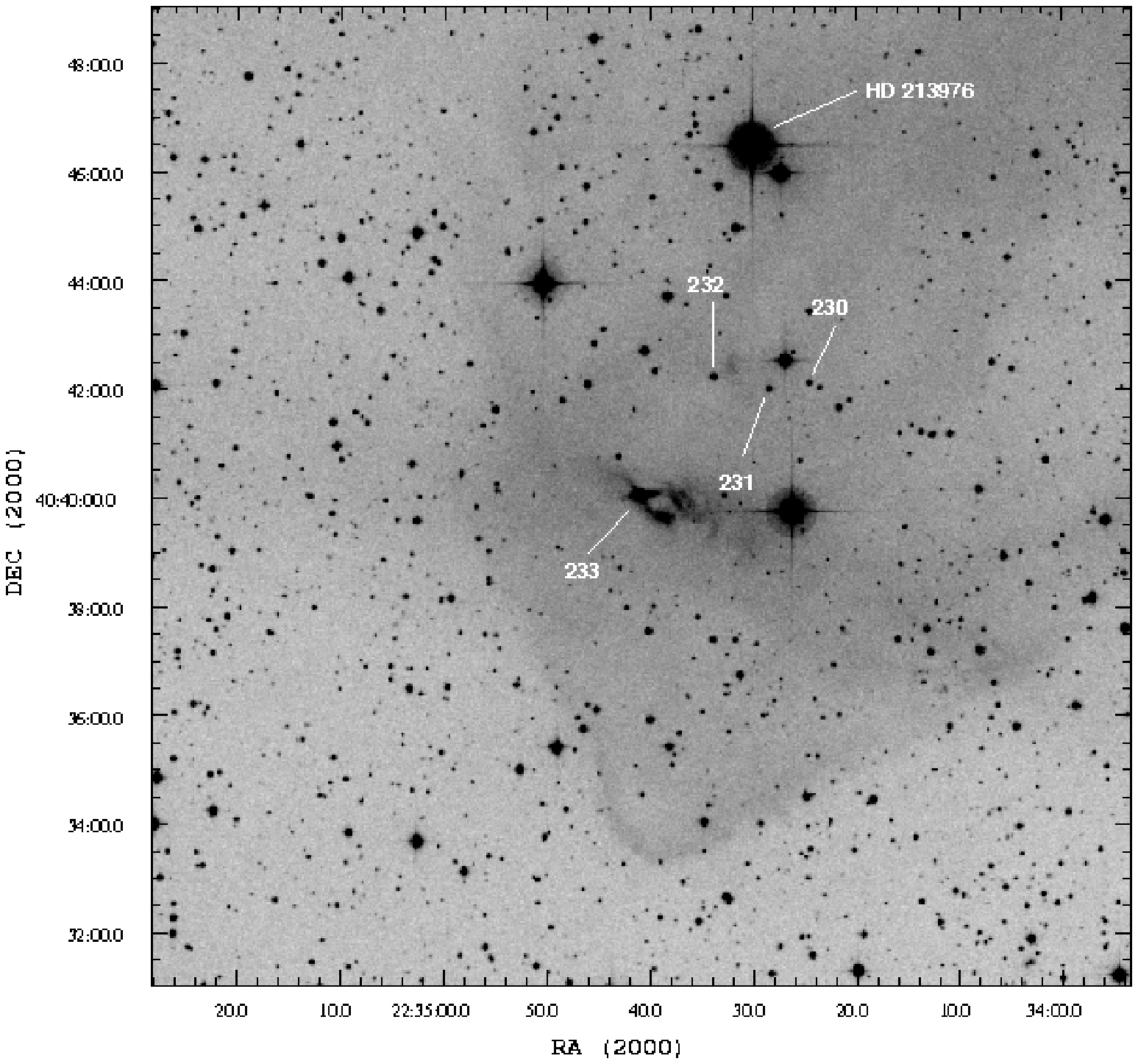}
 \caption{DSS-2 red image of the region around LkH$\alpha$\,233 
          and other emission-line stars, each labeled with its LkH$\alpha$ 
          number, and the luminous star HD\,213976.
     }
 \label{fig:lkha233dss2red}
\end{figure}

Stars associated with Condensation A include LkH$\alpha$\,233 (or V375\,Lac), 
a Herbig A4e star \citep{her04} showing H$\alpha$, [O\,I]\,6300~\AA, 
and [S\,II]\,6717~\AA\ emission
lines in the spectrum \citep{lee07}.  LkH$\alpha$\,233 was noticed by 
\citet{her60} to be an Ae/Be star associated with nebulosity that was 
later resolved by near-infrared speckle interferometry to be 
$\sim1000$~AU in size \citep{lei93}.  Given $m_{\rm V} = +13.8$ 
for LkH$\alpha$\,233, and assuming a luminosity class V (later known not 
to be valid), hence $M_{\rm V} = +2.3$, \citet{ode88} estimated a distance 
140--860~pc to LkH$\alpha$\,233, depending on the adopted value of 
optical extinction.  Fig.~\ref{fig:lkha233dss2red} shows the region 
around LKH$\alpha$\,233, including other fainter emission-line 
stars LkH$\alpha$\,230, LkH$\alpha$\,231, LkH$\alpha$\,232, 
and the luminous star HD\,213976 \citep{her60}.   

\begin{figure}[!ht] 
 \plotfiddle{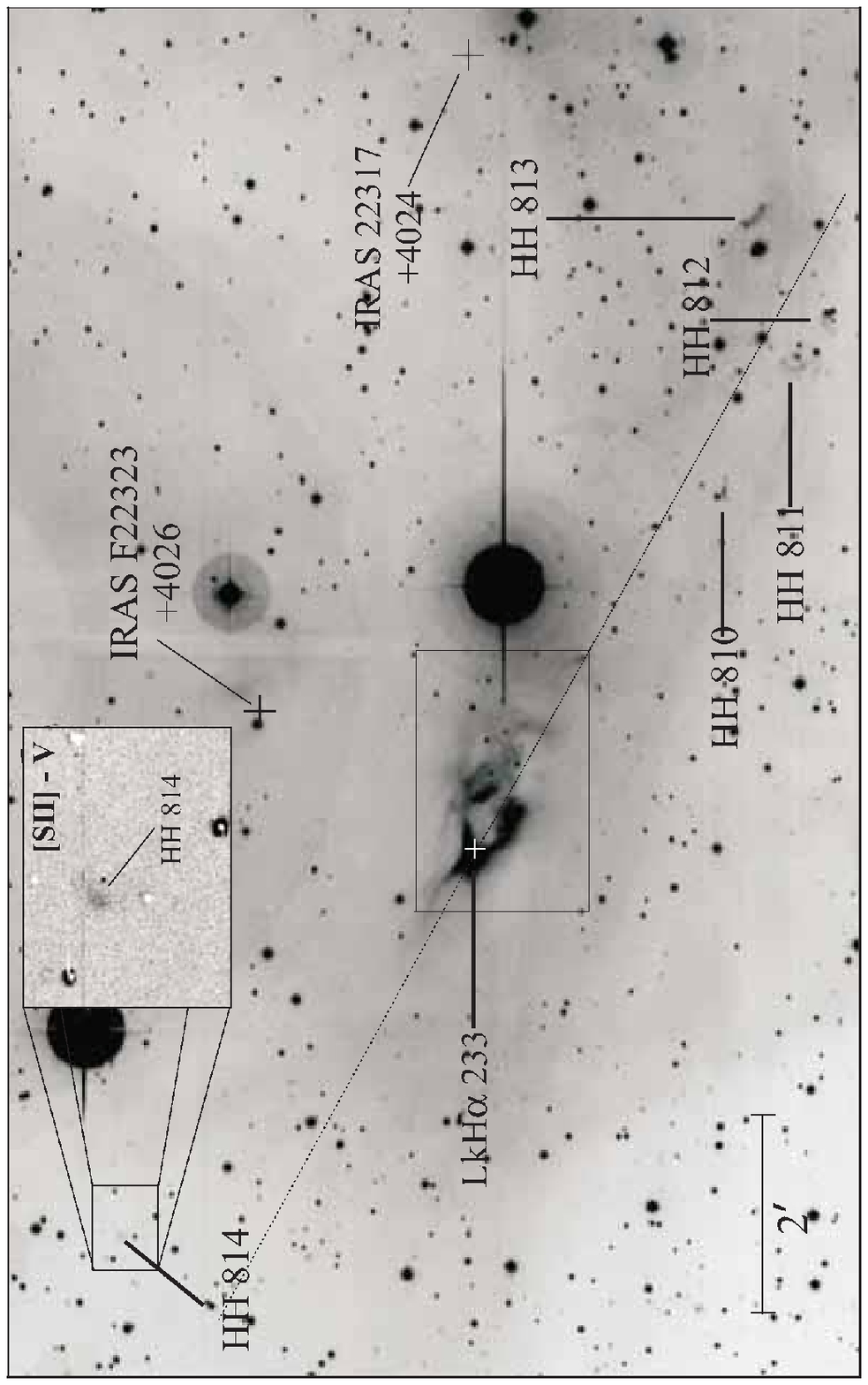}{8cm}{-90}{70}{70}{-300}{320}
 \caption{The [S\,II] image of the LkH$\alpha$\,233 region, taken from \citet{mcg04}.
          Herbig-Haro objects and IRAS
          sources are labeled.  The straight line depicts the major axis of the outflow at
          62\deg.  The inset shows the continuum subtracted ([S\,II]-V) image of
          HH\,814.  
     }
 \label{fig:lkha233}
\end{figure}

LkH$\alpha$\,233 is the exciting source of a 
series of bipolar Herbig-Haro objects \citep{cor98},
including HH\,398 and HH\,808 through HH\,814, that stretch a few
parsecs in length along roughly the direction of 65\deg/245\deg
\citep[see Fig.~\ref{fig:lkha233}]{mcg04}.  Note that \citet{mcg04}
adopted a distance of 880~pc to LkH$\alpha$\,233, apparently taken
from \citet{cal78} based on the inference that the B1.5~V star HD\,213976, with 
$m_V = 7.0$ and a distance modulus of 9.6, has a negligible extinction 
$A_V \sim 0.42$ \citep{asp85}, so should be in front of the dark cloud.  
In such a case, the cloud, and hence LkH$\alpha$\,233, should be at least 880~pc away.     
This inferred distance is however much farther than the recent $Hipparcos$
value of 370~pc \citep{dez99}, thus the linear dimensions of the 
LkH$\alpha$\,233 outflows derived by \citet{mcg04} should be a couple of 
times shorter---but still on parsec scales.  Most HH outflows are excited by low-mass
PMS stars, so the ones associated with LkH$\alpha$\,233 are among the
rarities to be related to intermediate-mass PMS stars
\citep{mcg04}.  Optical polarimetric imaging taken by \citet{asp85}
revealed a circumstellar disk roughly perpendicular to the outflows.
A recent high angular resolution imaging by Keck adaptive optics 
indicates that the bipolar jet of  LkH$\alpha$\,233, redshifted in 
position angle of 69\deg and blueshifted in 249\deg, 
is highly collimated, with an opening angle less than 10\deg, suggestive 
of an early accretion phase \citep{per07}.  This means the transition from 
highly collimated outflows typically seen in T Tauri stars,
to less collimated ones associated with massive young stars, must occur at 
a higher mass than the 4~M$_\odot$ estimated for LkH$\alpha$\,233 \citep{per07}.

Between 10\,Lac and LBN\,437 there is a group of PMS stars spanning
some 24$\arcmin$ (about 2.6~pc) across, most of which exhibit forbidden 
lines, indicative of youth \citep{lee07}.  LkH$\alpha$\,233 is
located near the edge of LBN\,437 and, being the exciting source of
Herbig-Haro objects, conceivably should be among the youngest.  There are
otherwise no CTTSs or Herbig Ae/Be stars known inside the cloud
\citep{lee07}.  The formation of this chain of young stars lying between 
10\,Lac and the LBN\,437 cloud complex might be triggered by the radiation-driven 
implosion mechanism (\citeauthor{ber89} \citeyear{ber89}, 
\citeauthor{ber90} \citeyear{ber90}, \citeauthor{hes05} 
\citeyear{hes05}), in which the UV photons from a luminous star evaporate
and compress a nearby molecular cloud.  As the result, the cloud is
shaped into a pillar, being illuminated to become a bright-rimmed cloud, and
star formation may take place at the surface layer of the cloud.


\subsection{GAL\,110$-$13} 

GAL\,110$-$13 is an isolated and elongated cloud \citep{whi49}. 
The CTTS BM\,And (RA = 23h37m38.5,
Decl = +48\deg 24\arcmin 12\arcsec, J2000) and three B-type stars
associated with the cloud, HD\,222142 (which illuminates the nebula
vdB\,158, \citeauthor{van57} \citeyear{van57}), HD\,222046, and
HD\,222086, all share common proper motions, suggesting a physical
group \citep{lee07}.  This cloud was not included in the study by
\citet{dez99}, but given its distance \citep[$\sim440$~pc,][]{ave69}, 
cloud radial velocity \citep[$\sim 8$~km~s$^{-1}$,][]{ode92}, 
proper motions \citep{lee07}, and association with young stars, it is likely 
a part of Lac\,OB1.

\citet{ode92} attributed the morphology and high star-forming efficiency (30\%) 
in GAL\,110$-$13 to compression by a recent cloud collision.  The cloud points to the 
central part of Lac\,OB1 where 10\,Lac is located, similar to LBN\,437 and other cloud 
filaments in the region (see Fig.~\ref{fig:lacob1co}).  Alternative to a cloud collision 
is shock interaction from a supernova in Lac\,OB1b which shaped GAL\,110$-$13 and 
prompted the formation of stars in the cloud.  Evidence in support of this 
supernova scenario comes from the B5V star HD\,201910, a possible runaway star from a binary
system in Lac\,OB1b when one of the component stars became a supernova
(\citeauthor{bla61} \citeyear{bla61}, \citeauthor{gie86} \citeyear{gie86}).

\begin{figure}[!ht] 
   \plotone{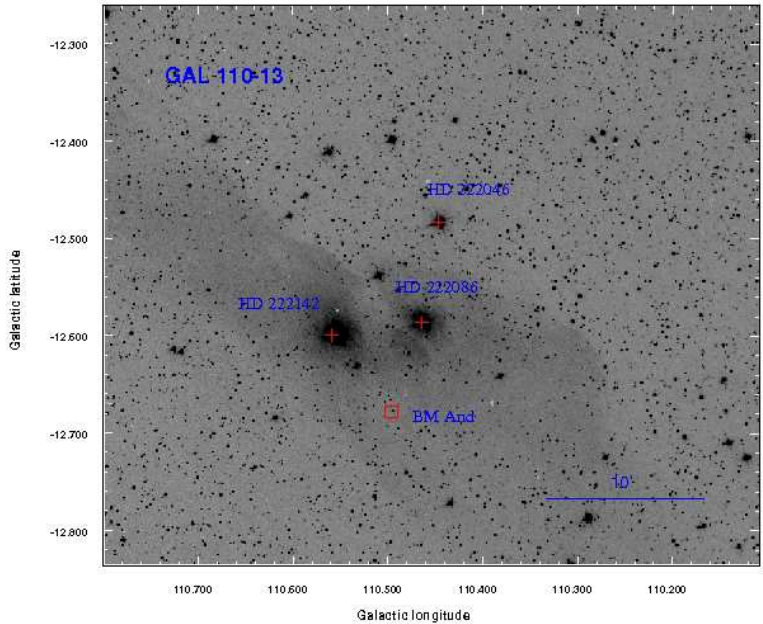}
 \caption{DSS blue image of GAL\,110$-$13, shown in Galactic coordinates.  BM\,And and three late-B
stars are marked.  
     }
 \label{fig:gal}
\end{figure}

%


\section{Star Formation History in Lac\,OB1}

\citet{bla58} and \citet{bla64,bla91} derived an expansion age of 16--25 Myr
for Lac\,OB1a and 12--16 Myr for Lac\,OB1b, on the basis of stellar proper
motions and radial velocities. The majority of the
Lac\,OB1 members indeed was thought to be an evolved population; e.g.,
\citet{her05} failed to find {\it bright} Herbig Ae/Be stars in the
region, and all the H$\alpha$ emission-line stars these authors
studied turned out to be classical Be stars, i.e., on the verge of
turning off the main sequence.  The kinematic ages of tens of Myr,
however, are much longer than the main sequence lifetime of
$\sim3.6$~Myr for 10\,Lac \citep{sch97} and the typical age of a few Myr for the
CTTSs in the region.  

\begin{figure}[!ht] 
 \plotone{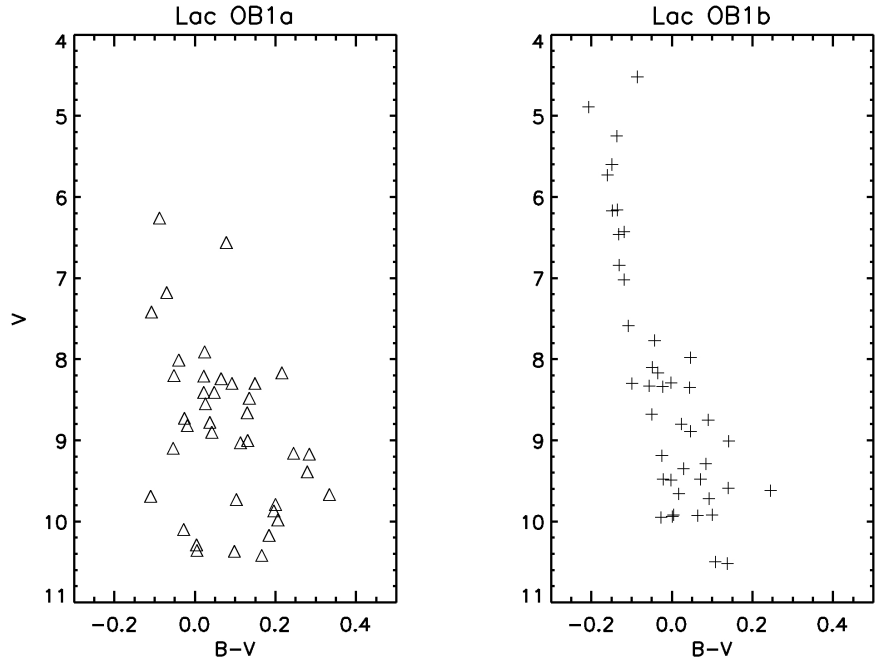}
\caption{Color-magnitude diagrams of the subgroups Lac\,OB1a and Lac\,OB1b reconstructed 
         from \citet{dez99}.  As in \citet{dez99}, stars having $\bv > 0.4$~mag are not shown.  
         The stars in Lac\,OB1b (pluses) form a clear main sequence, 
         whereas those in Lac\,OB1a (triangles) are much scattered.
         }
\label{fig:cmd}
\end{figure}

Star formation in Lac\,OB1 therefore appears not coeval,
with the latest episode occurring no more than a few Myr ago.
Kinematic ages of OB associations are often a factor of 2 less than
those derived photometrically based on stellar evolution models
\citep{gar94}.  Subgroups in an OB association may originate from a 
gravitationally unbound giant molecular cloud \citep{cla05}.
Likewise, members in a subgroup may be formed out of dispersing cloud
fragments, or as a consequence of triggered star formation by an
expanding ionization front.  Figure~\ref{fig:cmd} shows the
color-magnitude diagram for Lac\,OB1a and for Lac\,OB1b.  It is seen
that the stars in the subgroup Lac\,OB1b form a clear main sequence,
whereas those in Lac\,OB1a are much scattered.  \Citet{dez99}
suspected that Lac\,OB1a might not be a physical group.  In any case, care
should be exercised when doing photometric dating.  The scattering
could be attributed partly to the distance spread among members, as
Lac\,OB1a is nearby and occupies a large volume in space.  The ageing 
of Lac\,OB1b is evidenced by deficiency of H\,I gas around S\,126
where 10\,Lac and other luminous stars are located \citep{cap90}.
Lac\,OB1a, if it is a real association, should contain some PMS stars 
so represents a generation of stars younger than those in Lac\,OB1b.
Eventually the sequence of star formation reached GAL\,110$-$13, 
as we now witness.

Both LBN\,437 and GAL\,110$-$13 have low dust extinction, just like  
the bright-rimmed clouds in Ori\,OB1 \citep{lee05}, supporting the notion that 
they are remnant clouds \citep{sug91}.  Such a low density condition is unfavorable 
for spontaneous, global cloud collapse.  The ablation of molecular clouds also gives
rise to a seemingly high star-formation efficiency, e.g., 30\% for
GAL\,110$-$13 \citep{ode92}, to be compared with a few percent typical
in star-forming regions \citep{whi95}.  The cloud morphology, age sequence, and 
spatial distribution of young stars in the vicinity of clouds suggest 
triggered star formation by stellar radiation, supernova shocks or cloud collision.  
In particular, if GAL\,110$-$13 is indeed related to Lac\,OB1, the triggering 
appears to have far-reaching influence out to hundreds of parsecs.  
The Lac\,OB1 association, with much of the cloud
material already dissipated, is clearly ending its star-formation
activity, and stages an interesting case of the starbirth sequence in
an OB association.

\acknowledgements 
We thank the referee, Carlos Olano for very useful comments that much 
improved the quality of the article.  This work has made use of the 
NASA's Astrophysics Data System, and of the SIMBAD database, 
operated at CDS, Strasbourg, France.  The grant NSC-95-2745-M-008-002 
is acknowledged.



\begin{landscape}
\begin{table}[!ht]
\begin{center}
\caption{Kinematic Members of Lac\,OB1}
\smallskip
{\scriptsize 
\begin{tabular}{cccrrrrrrrll}
\tableline
\noalign{\smallskip}
 ID & RA & Decl  & \multicolumn{1}{c} {$V$} & \multicolumn{1}{c} {$\bv$} & \multicolumn{1}{c}{$\pi$} & 
    \multicolumn{1}{c}{$\mu_\alpha$ ~~~~~ $\mu_\delta$} & 
         \multicolumn{1}{c}{$J$} &  \multicolumn{1}{c}{$H$} &  \multicolumn{1}{c}{$K$} &      \multicolumn{1}{c}{Sp} & 
          \multicolumn{1}{c}{Remarks} \\
    & h~~~m~~~s & \deg~~~\arcmin~~~\arcsec & \multicolumn{1}{c}{mag} & \multicolumn{1}{c}{mag} & 
       \multicolumn{1}{c}{mas} & \multicolumn{1}{c}{mas/yr} & \multicolumn{1}{c}{mag} & \multicolumn{1}{c}{mag} & 
         \multicolumn{1}{c}{mag} & \multicolumn{1}{c}{Type} & \\
\noalign{\smallskip}
\tableline
\noalign{\smallskip}
HIP   108508 &        21    58 56.6  &        47    59   00  &     8.82  &    -0.019 &     1.32     ( 0.87)      &    -2.07 -3.38  &     8.66  &     8.70  &     8.69  &    B3V    &    MR\,Cyg, sp. binary                \\
HIP   109082 &        22   05  51.2  &        48    13   53  &     6.26  &    -0.088 &     1.66     ( 0.53)      &    -1.62 -4.33  &     6.36  &     6.50  &     6.48  &    B2V    &    V365\,Lac, sp. binary              \\
HIP   110476 &        22    22 41.4  &        42    57   04  &     9.29  &    0.084  &     1.91     ( 1.15)      &    -0.27 -6.32  &     8.91  &     8.87  &     8.76  &    B8     &    BD+42\,4370                        \\
HIP   110790 &        22    26 45.6  &        37    26   37  &     6.46  &    -0.132 &     1.69     ( 0.95)      &    -0.82 -5.20  &     6.70  &     6.84  &     6.87  &    B2V    &    double star                        \\
HIP   110835 &        22    27 17.3  &        44    22   46  &     9.92  &    0.004  &     2.68     ( 1.45)      &    -0.38 -4.18  &     9.86  &     9.86  &     9.87  &    B8     &    BD+43\,4205                        \\
HIP   110849 &        22    27 26.5  &        39    48   36  &     6.16  &    -0.136 &     2.39     ( 0.71)      &    -0.53 -6.09  &     6.41  &     6.54  &     6.53  &    B2V    &    HD\,212978, double/multiple        \\
HIP   110953 &        22    28 47.7  &        46    37   51  &     9.10  &    -0.054 &     2.25     ( 1.17)      &     0.17 -3.67  &     9.15  &     9.21  &     9.21  &    B9     &    HD\,213190                         \\
HIP   111080 &        22    30 12.3  &        44    26   18  &     8.89  &    0.046  &     1.52     ( 1.11)      &     0.99 -2.21  &     8.65  &     8.65  &     8.64  &    B9     &    HD\,213390                         \\
HIP   111104 &        22    30 29.3  &        43   07    24  &     4.52  &    -0.086 &     2.38     ( 0.64)      &    -2.05 -5.76  &     4.99  &     4.70  &     4.75  &    B2IV   &    HD\,213420, sp. binary             \\
HIP   111139 &        22    30 54.4  &        43    25   40  &     8.29  &    -0.002 &     1.19     ( 1.00)      &     0.63 -3.39  &     8.22  &     8.27  &     8.27  &    B9     &    HD\,213484                         \\
HIP   111308 &        22    32 58.6  &        37    34   32  &    10.52  &    0.138  &     5.45     ( 1.81)      &    -0.79 -3.39  &     8.67  &     8.21  &     8.13  &    B9     &    BD+36\,4868                        \\
HIP   111337 &        22    33 23.5  &        39    34   31  &     8.17  &    -0.035 &     2.35     ( 1.13)      &    -1.53 -4.29  &     8.18  &     8.23  &     8.26  &    B9V    &    HD\,213801, double/multiple        \\
HIP   111429 &        22    34 30.2  &        40    46   30  &     7.02  &    -0.119 &     3.45     ( 0.90)      &    -0.68 -3.45  &     7.19  &     7.27  &     7.29  &    B1.5V  &    HD\,213976, double                 \\
HIP   111491 &        22    35 18.1  &        43    40   52  &     8.33  &    -0.056 &     3.92     ( 0.93)      &    -0.10 -3.22  &     8.42  &     8.46  &     8.50  &    B8     &    HD\,214098                         \\
HIP   111546 &        22    35 52.3  &        39    38   04  &     5.73  &    -0.160 &     5.10     ( 1.79)      &     1.11 -4.39  &     5.78  &     5.85  &     5.70  &    B2Ve   &    HD\,214167                         \\
HIP   111576 &        22    36 16.7  &        40   05    20  &     8.30  &    -0.099 &     3.07     ( 1.01)      &     0.94 -3.22  &     8.50  &     8.58  &     8.59  &    B6IV     &    HD\,214243                         \\
HIP   111589 &        22    36 22.3  &        37    50   32  &     6.84  &    -0.131 &     2.92     ( 0.72)      &    -0.89 -5.34  &     7.12  &     7.23  &     7.25  &    B2V    &    HD\,214263                         \\
HIP   111683 &        22    37 28.7  &        39    26   20  &     7.59  &    -0.108 &     3.07     ( 0.79)      &    -0.34 -5.04  &     7.74  &     7.83  &     7.85  &    B3V    &    HD\,214432                         \\
HIP   111841 &        22    39 15.7  &        39   03    01  &     4.89  &    -0.207 &     3.08     ( 0.62)      &    -0.29 -5.70  &     5.30  &     5.44  &     5.50  &    O9V    &    10\,Lac, HD\,214680, double        \\
HIP   112031 &        22    41 28.7  &        40    13   32  &     5.25  &    -0.137 &     2.34     ( 0.62)      &    -0.75 -5.90  &     5.48  &     5.58  &     5.62  &    B2III  &    12\,Lac, HD\,214993, $\beta$ Cep var. \\
HIP   112144 &        22    42 55.4  &        37    48   10  &     6.43  &    -0.119 &     2.71     ( 0.79)      &    -1.12 -5.30  &     6.61  &     6.67  &     6.67  &    B1V    &    HD\,215191                         \\
HIP   112148 &        22    42 57.3  &        44    43   18  &     8.75  &    0.090  &     3.90     ( 1.34)      &    -2.82 -3.16  &     8.23  &     8.13  &     7.88  &    B5:ne  &    HD\,215227                         \\
HIP   112167 &        22    43 03.4  &        38    46   07  &     8.68  &    -0.050 &     1.80     ( 1.13)      &     0.67 -4.97  &     8.73  &     8.78  &     8.81  &    B8V    &    HD\,215211                         \\
HIP   112293 &        22    44 43.3  &        40    33   16  &     9.93  &    0.064  &     3.30     ( 1.61)      &    -1.66 -5.41  &     9.71  &     9.63  &     9.62  &    B8     &    BD+39\,4920                        \\
HIP   112906 &        22    51 50.2  &        39   08    42  &     9.49  &    -0.002 &     2.19     ( 1.41)      &     0.30 -5.12  &     9.44  &     9.49  &     9.48  &    B8     &    BD+38\,4883                        \\
HIP   113003 &        22    53 07.3  &        43   03    21  &     8.80  &    0.023  &     2.85     ( 1.12)      &    -0.18 -6.00  &     8.63  &     8.72  &     8.69  &    B9     &    HD\,216537                         \\
HIP   113110 &        22    54 21.2  &        43    31   43  &     7.77  &    -0.043 &     3.18     ( 0.85)      &     0.09 -4.79  &     7.76  &     7.87  &     7.85  &    B8V    &    HD\,216684                         \\
HIP   113226 &        22    55 47.1  &        43    33   33  &     7.98  &    0.046  &     4.01     ( 1.61)      &    -0.33 -4.29  &     7.67  &     7.61  &     7.46  &    B3V:n  &    V423\,Lac, HD\,216851              \\
HIP   113281 &        22    56 23.6  &        41    36   14  &     5.60  &    -0.149 &     2.71     ( 0.69)      &    -0.99 -4.25  &     5.88  &     6.01  &     6.03  &    B2IV   &    16\,Lac, HD\,216916, $\beta$ Cep var. \\
HIP   113371 &        22    57 40.7  &        39    18   32  &     6.17  &    -0.148 &     2.39     ( 0.66)      &     0.46 -5.13  &     6.46  &     6.61  &     6.64  &    B2IV/V &    HD\,217101                         \\
HIP   113469 &        22    58 45.7  &        43    50   20  &     7.18  &    -0.070 &     2.74     ( 0.72)      &     0.67 -5.75  &     7.25  &     7.31  &     7.33  &    B2:V   &    HD\,217227                         \\
HIP   113835 &        23   03  08.3  &        49    35   10  &     9.69  &    -0.110 &     3.14     ( 1.37)      &    -2.33 -1.53  &     9.64  &     9.68  &     9.67  &    B8     &    BD+48\,3916                        \\
\noalign{\smallskip}
\tableline
\end{tabular}
}
\end{center}
\end{table}

\setcounter{table}{0}

\begin{table}[!ht]
\caption{Kinematic Members of Lac\,OB1 (continued)}
\smallskip
\begin{center}
{\scriptsize 
 \begin{tabular}{cccrrrrrrrll}
\tableline
\noalign{\smallskip}
  ID & RA & Decl  & \multicolumn{1}{c} {$V$} & \multicolumn{1}{c} {$\bv$} & \multicolumn{1}{c}{$\pi$} &
    \multicolumn{1}{c}{$\mu_\alpha$ ~~~~~ $\mu_\delta$} &
         \multicolumn{1}{c}{$J$} &  \multicolumn{1}{c}{$H$} &  \multicolumn{1}{c}{$K$} &     \multicolumn{1}{c}{Sp} & 
           \multicolumn{1}{c}{Remarks} \\
    & h~~~m~~~s & \deg~~~\arcmin~~~\arcsec & \multicolumn{1}{c}{mag} & \multicolumn{1}{c}{mag} &
       \multicolumn{1}{c}{mas} & \multicolumn{1}{c}{mas/yr} & \multicolumn{1}{c}{mag} & \multicolumn{1}{c}{mag} &
         \multicolumn{1}{c}{mag} & \multicolumn{1}{c}{Type} & \\
\noalign{\smallskip}
\tableline
\noalign{\smallskip}
%
HIP   114097 &        23   06  32.2  &        51   04    38  &     7.42  &    -0.108 &     2.41     ( 0.67)      &    -0.28 -4.13  &     7.63  &     7.73  &     7.76  &    B2V    &    HD\,218344                         \\
HIP   114106 &        23   06  37.1  &        42    39   27  &     8.01  &    -0.040 &     4.57     ( 0.87)      &    -1.81 -4.54  &     8.06  &     8.10  &     8.16  &    B9     &    V380\,And, HD\,218326              \\
HIP   115067 &        23    18 23.6  &        47    15   42  &     8.66  &    0.130  &     3.77     ( 1.52)      &    -1.85 -2.62  &     8.54  &     8.57  &     8.64  &    B9II   &    HD\,219813, double                 \\
HIP   115334 &        23    21 38.7  &        47    21   04  &     8.41  &    0.048  &     3.28     ( 0.90)      &    -0.93 -5.50  &     8.21  &     8.17  &     8.14  &    B9     &    HD\,220210                         \\
HIP   106656 &        21    36 11.4  &        44    25   38  &     8.90  &    0.042  &     2.53     ( 0.95)      &     1.02 -3.46  &     8.77  &     8.80  &     8.79  &    A0     &    HD\,205742                         \\
HIP   108841 &        22   02  54.6  &        39    33   46  &     8.21  &    0.022  &     1.71     ( 1.50)      &    -1.07 -2.91  &     8.08  &     8.09  &     8.10  &    A0     &    HD\,209483, double                 \\
HIP   108933 &        22   04  06.7  &        44    20   42  &     6.56  &    0.078  &     5.72     ( 0.67)      &    -0.53 -2.90  &     6.29  &     6.31  &     6.27  &    A2     &    HD\,209679                         \\
HIP   110033 &        22    17 12.0  &        40    58   05  &     9.48  &    0.071  &     1.80     ( 1.25)      &     0.96 -3.41  &     9.36  &     9.39  &     9.38  &    A0     &    BD+40\,4771                        \\
HIP   110373 &        22    21 21.1  &        41    47   48  &     8.34  &    -0.023 &     3.25     ( 1.11)      &    -2.23 -5.81  &     8.36  &     8.42  &     8.42  &    A0     &    HD\,212153, double/multiple        \\
HIP   110448 &        22    22 17.9  &        48    50   25  &     8.39  &    1.172  &     1.89     ( 0.79)      &    -2.02 -2.48  &     6.30  &     5.79  &     5.63  &    K0     &    BD+48\,3697                        \\
HIP   110473 &        22    22 38.5  &        47    37   56  &     9.98  &    0.206  &     2.33     ( 1.92)      &     0.12 -4.51  &     9.62  &     9.55  &     9.51  &    A0     &    BD+46\,3676                        \\
HIP   110664 &        22    25 06.0  &        44    32   19  &     8.10  &    -0.049 &     1.87     ( 0.82)      &    -1.85 -5.22  &     8.19  &     8.28  &     8.26  &    A0     &    HD\,212668                         \\
HIP   110700 &        22    25 43.7  &        38    49   26  &     9.48  &    -0.022 &     5.20     ( 1.37)      &    -0.16 -6.18  &     9.45  &     9.51  &     9.51  &    A0     &    HD\,212732                         \\
HIP   110804 &        22    26 58.3  &        46   01    49  &    10.29  &    0.004  &     2.79     ( 1.32)      &    -2.01 -3.57  &    10.12  &    10.15  &    10.17  &    A0     &    BD+45\,3940                        \\
HIP   110929 &        22    28 29.3  &        48    32   34  &     7.84  &    1.800  &     2.09     ( 1.02)      &    -1.82 -4.17  &     5.26  &     4.62  &     4.39  &    K0III: &    HD\,213141, double                 \\
HIP   111022 &        22    29 31.8  &        47    42   25  &     4.34  &    1.679  &     2.80     ( 0.50)      &    -0.60 -3.37  &     1.32  &     0.41  &     0.27  &    M0II:  &    5\,Lac,  HD\,213310/213311         \\
HIP   111038 &        22    29 42.7  &        40    55   19  &     9.62  &    0.245  &     1.87     ( 1.57)      &     0.52 -2.45  &     9.04  &     8.98  &     8.94  &    A5     &    BD+40\,4831                        \\
HIP   111055 &        22    29 52.6  &        45    44   41  &     7.80  &    1.334  &     1.73     ( 0.87)      &     1.41 -2.34  &     5.42  &     4.82  &     4.64  &    K2     &    HD\,213354                         \\
HIP   111207 &        22    31 45.3  &        43    16   52  &     9.95  &    -0.027 &     1.71     ( 1.50)      &    -2.12 -3.00  &     9.87  &     9.88  &     9.88  &    A0     &    BD+42\,4429                        \\
HIP   111292 &        22    32 43.1  &        46    16   21  &    10.10  &    -0.028 &     2.19     ( 1.50)      &     1.40 -5.29  &    10.01  &    10.05  &    10.08  &    Ap     &    HD\,213732                         \\
HIP   111329 &        22    33 19.9  &        42    23   42  &     9.19  &    -0.025 &     3.20     ( 1.26)      &     0.48 -3.63  &     9.14  &     9.17  &     9.14  &    A0     &    HD\,213800                         \\
HIP   111340 &        22    33 25.1  &        46    51   27  &     9.39  &    0.279  &     1.38     ( 1.21)      &    -1.02 -4.08  &     8.69  &     8.57  &     8.53  &    A2     &    HD\,213833                         \\
HIP   111375 &        22    33 48.1  &        41    40   28  &     9.94  &    0.001  &     2.07     ( 2.51)      &     0.41 -5.11  &     9.94  &    10.01  &     9.99  &    A0     &    BD+40\,4852                        \\
HIP   111552 &        22    35 54.5  &        43    41   26  &     9.66  &    0.017  &     2.36     ( 1.42)      &    -0.58 -4.08  &     9.57  &     9.59  &     9.58  &    A0     &    HD\,214179                         \\
HIP   111591 &        22    36 25.0  &        46    55   39  &    10.37  &    0.098  &     2.15     ( 1.67)      &    -0.31 -3.88  &    10.18  &    10.20  &    10.19  &    A0     &    HD\,214311                         \\
HIP   111762 &        22    38 22.2  &        52    22   06  &     9.98  &    1.363  &     8.98     ( 3.27)      &    -0.85 -5.59  &     7.26  &     6.53  &     6.34  &     -     &    BD+51\,3434, double/multiple       \\
HIP   111814 &        22    38 54.9  &        36    55   42  &     9.72  &    0.092  &     4.06     ( 1.41)      &    -2.83 -4.82  &     9.42  &     9.42  &     9.41  &    A2     &    BD+36\,4896                        \\
HIP   111916 &        22    40 12.5  &        38    58   25  &     9.59  &    0.140  &     2.79     ( 1.47)      &    -2.07 -6.15  &     9.27  &     9.25  &     9.22  &    A2     &    BD+38\,4834                        \\
HIP   112016 &        22    41 22.9  &        50   05    33  &     7.91  &    0.024  &     2.38     ( 0.74)      &     0.10 -2.74  &     7.78  &     7.81  &     7.80  &    A0     &    HD\,215025                         \\
HIP   112017 &        22    41 23.7  &        41   02    16  &     9.35  &    0.029  &     2.27     ( 1.33)      &    -1.04 -5.40  &     9.29  &     9.28  &     9.28  &    A2     &    HD\,214977                         \\
HIP   112182 &        22    43 15.2  &        43    46   25  &    10.50  &    0.108  &     3.77     ( 1.64)      &    -1.02 -5.11  &    10.14  &    10.15  &    10.10  &    A0     &    HD\,215271                         \\
\noalign{\smallskip}
\tableline
\end{tabular}
}
\end{center}
\end{table}

\setcounter{table}{0}

\begin{table}[!ht]
\caption{Kinematic Members of Lac\,OB1 (continued) }
\smallskip
\begin{center}
{\scriptsize 
 \begin{tabular}{cccrrrrrrrll}
\tableline
\noalign{\smallskip}
  ID & RA & Decl  & \multicolumn{1}{c} {$V$} & \multicolumn{1}{c} {$\bv$} & \multicolumn{1}{c}{$\pi$} &
    \multicolumn{1}{c}{$\mu_\alpha$ ~~~~~ $\mu_\delta$} &
         \multicolumn{1}{c}{$J$} &  \multicolumn{1}{c}{$H$} &  \multicolumn{1}{c}{$K$} &      \multicolumn{1}{c}{Sp} & 
           \multicolumn{1}{c}{Remarks} \\
    & h~~~m~~~s & \deg~~~\arcmin~~~\arcsec & \multicolumn{1}{c}{mag} & \multicolumn{1}{c}{mag} &
       \multicolumn{1}{c}{mas} & \multicolumn{1}{c}{mas/yr} & \multicolumn{1}{c}{mag} & \multicolumn{1}{c}{mag} &
         \multicolumn{1}{c}{mag}  & \multicolumn{1}{c}{Type} & \\
\noalign{\smallskip}
\tableline
\noalign{\smallskip}
%
HIP   112212 &        22    43 35.3  &        32    49   19  &     7.27  &    1.628  &     1.92     ( 0.81)      &    -1.00 -3.11  &     4.14  &     3.31  &     2.90  &    M0III  &    QU\,Peg, HD\,215290                \\
HIP   112213 &        22    43 36.7  &        40    23   06  &     9.92  &    0.100  &     4.63     ( 1.67)      &    -2.60 -5.55  &     9.54  &     9.45  &     9.43  &    A2     &    BD+39\,4917                        \\
HIP   112639 &        22    48 47.2  &        46    22   11  &     9.03  &    0.113  &     2.31     ( 1.18)      &     0.61 -1.98  &     8.75  &     8.73  &     8.72  &    A0     &    HD\,216037                         \\
HIP   112700 &        22    49 21.6  &        45    53   50  &     8.48  &    0.135  &     2.02     ( 1.04)      &     0.72 -2.33  &     8.13  &     8.13  &     8.10  &    A0     &    HD\,216107                         \\
HIP   112710 &        22    49 29.4  &        45    46   59  &     9.87  &    0.196  &     2.13     ( 1.55)      &     0.40 -3.25  &     9.45  &     9.44  &     9.44  &    A2     &    HD\,216117                         \\
HIP   112805 &        22    50 40.5  &        51   06    58  &     8.30  &    0.149  &     2.00     ( 1.70)      &    -1.12 -6.07  &     8.06  &     8.09  &     8.05  &    A0     &    HD\,216255, double/multiple        \\
HIP   113145 &        22    54 43.9  &        42    30   39  &     7.83  &    0.520  &     3.74     ( 0.84)      &     1.50 -2.53  &     6.54  &     6.37  &     6.25  &    A2     &    HD\,216733                         \\
HIP   113187 &        22    55 13.8  &        46    22   20  &     8.30  &    0.092  &     1.91     ( 0.88)      &     1.34 -2.92  &     8.02  &     8.02  &     8.03  &    A0     &    HD\,216797                         \\
HIP   113188 &        22    55 14.1  &        49    58   42  &     8.78  &    0.037  &     1.59     ( 1.08)      &    -1.97 -3.50  &     8.64  &     8.68  &     8.67  &    A2     &    HD\,216795                         \\
HIP   113208 &        22    55 31.5  &        43    17   36  &     8.35  &    0.044  &     2.99     ( 1.03)      &     0.01 -3.65  &     8.20  &     8.21  &     8.26  &    A2     &    HD\,216815                         \\
HIP   113237 &        22    55 52.9  &        41    58   32  &     8.13  &    1.312  &     2.21     ( 0.96)      &    -0.68 -4.78  &     5.66  &     5.00  &     4.84  &    K2     &    HD\,216853                         \\
HIP   113288 &        22    56 26.0  &        49    44   01  &     4.99  &    1.778  &     1.74     ( 0.58)      &     0.05 -2.87  &     1.79  &     0.96  &     0.72  &    K5Ib:  &    V424\,Lac, HD\,216946              \\
HIP   113411 &        22    58 06.7  &        41    56   04  &     9.01  &    0.141  &     2.96     ( 1.78)      &     0.41 -3.98  &     9.06  &     9.12  &     9.07  &    A2     &    HD\,217161, double                 \\
HIP   113474 &        22    58 49.6  &        46    19   38  &     8.24  &    0.065  &     1.83     ( 0.89)      &    -0.01 -3.92  &     7.99  &     8.05  &     8.06  &    A0     &    HD\,217262                         \\
HIP   113731 &        23   01  58.4  &        47   01    00  &     8.41  &    0.022  &     1.81     ( 1.03)      &    -1.47 -2.48  &     8.21  &     8.23  &     8.24  &    A2     &    HD\,217713                         \\
HIP   113950 &        23   04  35.5  &        44    13   10  &    10.36  &    0.005  &     4.35     ( 2.67)      &    -0.06 -3.28  &     9.99  &    10.00  &     9.95  &    A0     &    BD+43\,4383                        \\
HIP   114134 &        23   06  54.0  &        44    19   26  &     8.55  &    0.026  &     2.58     ( 1.00)      &     0.92 -4.19  &     8.46  &     8.53  &     8.52  &    A0     &    HD\,218364                         \\
HIP   114153 &        23   07  05.6  &        46   07    47  &    10.17  &    0.184  &     2.13     ( 1.34)      &    -0.22 -4.95  &     9.82  &     9.78  &     9.79  &    A0     &    BD+45\,4144                        \\
HIP   114441 &        23    10 37.7  &        46    22   40  &     8.20  &    -0.052 &     2.22     ( 0.90)      &     0.27 -3.76  &     8.24  &     8.27  &     8.30  &    A0     &    HD\,218844                         \\
HIP   114554 &        23    12 15.0  &        38    46   59  &     9.16  &    0.245  &     4.17     ( 1.46)      &    -2.40 -5.09  &     8.64  &     8.57  &     8.55  &    A5     &    HD\,219016                         \\
HIP   114593 &        23    12 52.4  &        48    17   01  &     9.73  &    0.103  &     2.25     ( 1.38)      &     0.04 -5.32  &     9.57  &     9.60  &     9.56  &    A0     &    BD+47\,4075                        \\
HIP   114625 &        23    13 15.1  &        45    50   25  &    10.42  &    0.166  &     2.41     ( 1.83)      &    -1.81 -3.19  &    10.17  &    10.20  &    10.18  &    A2     &    BD+45\,4171                        \\
HIP   114642 &        23    13 26.7  &        35    45   43  &     8.65  &    1.100  &     4.72     ( 1.26)      &    -1.06 -6.02  &     6.79  &     6.29  &     6.19  &    K0     &    BD+34\,4870                        \\
HIP   114890 &        23    16 19.0  &        50   01    41  &     9.17  &    0.284  &     2.48     ( 1.04)      &    -0.73 -2.65  &     8.69  &     8.65  &     8.58  &    A0     &    HD\,219574                         \\
HIP   114909 &        23    16 31.1  &        36    50   13  &     9.67  &    0.334  &     2.07     ( 1.42)      &     1.46 -3.79  &     8.94  &     8.87  &     8.79  &    F0     &    BD+36\,5034                        \\
HIP   115441 &        23    23 00.4  &        38    59   57  &     9.00  &    0.131  &     2.36     ( 1.14)      &    -0.89 -3.50  &     8.68  &     8.67  &     8.65  &    A2     &    BD+38\,4988                        \\
HIP   116088 &        23    31 23.3  &        43    22   24  &     8.17  &    0.216  &     6.96     ( 0.94)      &    -1.97 -5.62  &     7.81  &     7.75  &     7.73  &    A0     &    HD\,221379                         \\
HIP   116135 &        23    31 53.1  &        47    32   52  &     9.79  &    0.200  &     2.24     ( 1.48)      &     1.70 -2.58  &     9.47  &     9.48  &     9.41  &    A2     &    BD+46\,4070                        \\
HIP   116411 &        23    35 24.2  &        36    44   54  &     9.19  &    1.628  &     3.19     ( 1.19)      &     1.69 -2.50  &     4.85  &     3.88  &     3.59  &    M2     &    V391\,And, BD+35\,5056             \\
HIP   116457 &        23    35 50.9  &        47    25   39  &     8.73  &    1.071  &     1.52     ( 1.11)      &     1.54 -2.04  &     6.83  &     6.33  &     6.22  &    K0     &    BD+46\,4089                        \\
HIP   116522 &        23    36 51.8  &        43    18   32  &     7.81  &    1.198  &     2.02     ( 0.88)      &     0.66 -5.07  &     5.79  &     5.18  &     5.09  &    K2     &    HD\,222018                         \\
HIP   116540 &        23    37 05.9  &        46    17   15  &     8.73  &    -0.026 &     2.17     ( 1.13)      &     1.06 -3.15  &     8.68  &     8.72  &     8.75  &    A0     &    HD\,222064                         \\
HIP   116681 &        23    38 45.1  &        35    46   21  &     9.74  &    2.100  &     5.50     ( 2.83)      &     1.58 -3.66  &     4.83  &     3.94  &     3.10  &    C      &    ST\,And, HD\,222241, carbon star   \\
\noalign{\smallskip}
\tableline
\end{tabular}
}
\end{center}
\end{table}
\end{landscape}

\end{document}